%% file: main.tex
\documentclass[conference,compsoc]{IEEEtran}
\ifCLASSINFOpdf
\else
\fi
\AtBeginDocument{%
  \providecommand\BibTeX{{%
    \normalfont B\kern-0.5em{\scshape i\kern-0.25em b}\kern-0.8em\TeX}}}
\usepackage{cite}
\usepackage{tikz}
\usepackage{amsmath}
\usepackage{booktabs}
\usepackage[flushleft]{threeparttable}
\usepackage{tabularx}
\usepackage{multirow}
\usepackage{multicol}
\usepackage{caption}
\usepackage{siunitx}
\usepackage{colortbl}
\usepackage{xcolor}
\usepackage{placeins}
\usepackage{makecell}
\usepackage{orcidlink}
\usepackage{hyperref}
\usepackage{amsmath}
\usepackage{amssymb}
\usepackage[T1]{fontenc}
\usepackage{color, colortbl}
\definecolor{LightGray}{gray}{0.95}
\usepackage{adjustbox}
\usepackage{pifont}
\newcommand{\cmark}{\ding{51}}%
\newcommand{\xmark}{\ding{55}}%
\usepackage[%
    autopunct=true,
]{csquotes} 

\SetBlockThreshold{3}
\usepackage[inline,shortlabels]{enumitem} 
\usepackage{tikz}
\usetikzlibrary{positioning, arrows, calc, backgrounds}
\usepackage{tikz}
\usepackage[tikz,usetwoside=true]{mdframed}
\usetikzlibrary{calc,arrows}
\tikzset{
    summary arrow/.style={%
        line width=1pt,
        draw=gray!40,
        rounded corners=1ex,
    },
    summary head/.style={
        fill=white,
        font=\bfseries\sffamily,
        text=gray!80,
        anchor=base west,
    },
}
\mdfdefinestyle{summary}{%
    singleextra={%
        \path let \p1=(P), \p2=(O) in (\x2,\y1) coordinate (Q);
        \path let \p1=(P), \p2=(O) in (\x1,\y2) coordinate (R);
        \path let \p1=(Q), \p2=(O) in (\x1,{(\y1-\y2)/2}) coordinate (M);
        \path [summary arrow] (R) -| (M) |- (P);
        \node [summary head] at ($(Q)+(1ex,-1.5pt)$) {\frametitle};
    },
    firstextra={%
        \path let \p1=(P), \p2=(O) in (\x2,\y1) coordinate (Q);
        \path [summary arrow,latex-] (O) |- (P);
        \node [summary head] at ($(Q)+(1ex,-1.5pt)$) {\frametitle};
    },
    secondextra={%
        \path let \p1=(P), \p2=(O) in (\x2,\y1) coordinate (Q);
        \path let \p1=(P), \p2=(O) in (\x1,\y2) coordinate (R);
        \path [summary arrow] (R) -| (Q);
    },
    middleextra={%
        \path let \p1=(P), \p2=(O) in (\x2,\y1) coordinate (Q);
        \path [summary arrow] (O) -- (Q);
    },
    middlelinewidth=1.5em,middlelinecolor=white,
    hidealllines=true,topline=true,
    innertopmargin=0pt,
    innerbottommargin=1em,
    innerrightmargin=0pt,
    innerleftmargin=1.5ex,
    skipabove=0.5\baselineskip,
    skipbelow=0.3\baselineskip,
}
\newenvironment{summary}[1]{
\begingroup
\def\frametitle{\small #1}
\begin{mdframed}[style=summary]
}{
\end{mdframed}
\endgroup
}

\DeclareQuoteStyle[american]{english} 
        {\itshape\textquotedblleft}
        [\textquotedblleft]
        {\textquotedblright}
        [0.05em]
        {\textquoteleft}
        {\textquoteright}

\newcounter{taskcount}
\newcommand*{\thetask}{\arabic{taskcount}}
\newcommand*{\deftask}[1]{%
  \leavevmode
  \refstepcounter{taskcount}%
  \label{#1}%
  Task \thetask%
}

\newcommand*{\task}[1]{Task~\ref{#1}}
\newcommand*{\tasks}[1]{\ref{#1}}
\ifdefined\DONE
\newcommand{\todo}[1]{}
\newcommand{\dk}[1]{}
\else%
\newcommand{\todo}[1]{\textcolor{red}{TODO:~#1}}
\newcommand{\dk}[1]{\textcolor{cyan}{TODO(DK):~#1}}

\fi%
\begin{document}



\title{{\it``Sorry for bugging you so much.''}\\ Exploring Developers' Behavior Towards Privacy-Compliant Implementation}









\author{
    \IEEEauthorblockN{Stefan Albert Horstmann\orcidlink{0000-0002-4053-0706}\IEEEauthorrefmark{1}, Sandy Hong\orcidlink{0009-0002-9206-6893}\IEEEauthorrefmark{1}, David Klein\orcidlink{0000-0001-8468-8516}\IEEEauthorrefmark{2}, Raphael Serafini\orcidlink{0000-0002-9137-5072}\IEEEauthorrefmark{3},\\ Martin Degeling\orcidlink{0000-0001-7048-781X}\IEEEauthorrefmark{4}, Martin Johns\orcidlink{0000-0003-2574-5060}\IEEEauthorrefmark{2}, Veelasha Moonsamy\orcidlink{0000-0001-6296-2182}\IEEEauthorrefmark{1}, Alena Naiakshina\orcidlink{0009-0008-1843-2027}\IEEEauthorrefmark{3}}
    \IEEEauthorblockA{\IEEEauthorrefmark{1}Ruhr University Bochum
    \{stefan-albert.horstmann, sandy.hong, veelasha.moonsamy\}@rub.de}
    \IEEEauthorblockA{\IEEEauthorrefmark{2}Technische Universität Braunschweig
    \{david.klein, m.johns\}@tu-braunschweig.de}
    \IEEEauthorblockA{\IEEEauthorrefmark{3}University of Cologne
    \{raphael.serafini, alena.naiakshina\}@uni-koeln.de}
    \IEEEauthorblockA{\IEEEauthorrefmark{4}Independent
    \{martin\}@degeling.com}
}


\maketitle
\begin{abstract}

While protecting user data is essential, software developers often fail to fulfill privacy requirements. However, the reasons why they struggle with privacy-compliant implementation remain unclear. Is it due to a lack of knowledge, or is it because of insufficient support?
To provide foundational insights in this field, we conducted a qualitative 5-hour programming study with 30 professional software developers implementing 3 privacy-sensitive programming tasks that were designed with GDPR compliance in mind. To explore if and how developers implement privacy requirements, participants were divided into 3 groups: control, privacy prompted, and privacy expert-supported.  
After task completion, we conducted follow-up interviews.
Alarmingly, almost all participants submitted non-GDPR-compliant solutions (79/90). In particular, none of the 3 tasks were solved privacy-compliant by all 30 participants, with the non-prompted group having the lowest number of 3 out of 30 privacy-compliant solution attempts. Privacy prompting and expert support only slightly improved participants' submissions, with 6/30 and 8/30 privacy-compliant attempts, respectively. 
In fact, all participants reported severe issues addressing common privacy requirements such as purpose limitation, user consent, or data minimization.
Counterintuitively, although most developers exhibited minimal confidence in their solutions, they rarely sought online assistance or contacted the privacy expert, with only 4 out of 10 expert-supported participants explicitly asking for compliance confirmation.
Instead, participants often relied on existing implementations and focused on implementing functionality and security first. 

\end{abstract}





\input{content/01_intro}
\input{content/02_relatedwork}
\input{content/03_methodology}
\input{content/04_results}

\input{content/05_discussion}

\input{content/06_conclusion}
\input{content/07_availability}
\input{content/08_acknowledgement}
\bibliographystyle{IEEEtran}
\bibliography{./main}{}

\appendices
\input{content/a5_interview}

\input{content/a3_screening}
\input{content/a4_postsurvey}
\input{content/a10_metareview}

\end{document}

%% file: content/01_intro.tex
\section{Introduction}\label{sec:introduction}
Several regulations have been passed to reign in tech companies' data collection practices and empower users to regain control over their personal data.
The best-known ones are Japan’s Act on the Protection of Personal Information (APPI) from 2003, South Africa’s Protection of Personal Information Act (POPI Act) from 2013, the General Data Protection Regulation (GDPR)~\cite{GDPR} (2018) by the European Union, and the California Consumer Privacy Act CCPA~\cite{CCPA} (2018) by the state of California.
Consequently, user privacy has become an essential requirement that software development companies must accomplish.
Severe fines~\cite{SaeTheUrb+22}, fear of bad publicity~\cite{CambrideAnalytica}, and customer expectations~\cite{horstmann2024those} all lead to an increased focus on developing privacy-preserving software products.
Still, since the introduction of e.g., the GDPR in 2018, more than 5 billion euros have been issued for violations~\cite{enforcementtracker}.
While larger companies often have access to legal experts to support software developers, their communication with software development teams tasked with fulfilling these requirements is often limited, suggesting a ``communication gap''~\cite{horstmann2024those}. Although smaller companies are hit harder by fines for privacy non-compliance, they are often unable to hire privacy experts due to limited budget restrictions. As a result, fulfilling privacy requirements has become a task handled solely by the software development teams. 

However, past research showed that this has not (yet) translated into the actual practices of individual developers (e.g.,~\cite{ SaeTheUrb+22, KocAltJoh23, NguBacMarSto21, NguBacSto22, tahaei2022embedding, kekulluouglu2023we, prybylo2024evaluating}), raising the question: are developers unaware of privacy requirements, or is the task of fulfilling them lacking support?
To address this research gap, we conducted a lab programming study with \num{30} professional software developers implementing requirements for a health application, lasting 3 hours and 23 minutes on average.
We asked participants to complete 4 programming tasks, with 3 of them requiring a GDPR-compliant implementation: (i) a warm-up task consisting of fixing an application's backup function, (ii) processing a deletion request from one of the application's users, (iii) implementing a search function for doctor users of the application, and lastly, (iv) implementing an e-mail function to send user advertisements and obtain user consent for a privacy-compliant solution.
With this work, we explored the following research questions:
\begin{description}
\item[\textbf{RQ1:}]``\textit{How do developers implement privacy measures in their software?}''
\item[\textbf{RQ2:}]``\textit{How do developers behave if they require information on privacy during implementation?}''
\item[\textbf{RQ3:}]``\textit{What issues do developers encounter during privacy measures implementation?}''
\end{description}

To explore developers' privacy awareness and address the communication gap~\cite{horstmann2024those}, we separated the participants into three groups: (i)~a control group without mentioning privacy at all, (ii)~a group prompted on privacy-compliant implementation, and (iii)~a chatting group having access to a privacy expert. 
To further understand participants' implementation challenges with privacy requirements during the study, we conducted follow-up interviews. 

With this work, we contribute foundational insights into a highly complex problem. 
Our findings suggest a lack of privacy awareness, the perception that privacy requirements are overly complex to implement, and a reluctance to consult privacy experts, as developers often fear being a burden or inconveniencing them.
Participants cited a lack of knowledge and experience and relied on preexisting implementations. 
Still, most interactions with experts were limited to confirming compliance rather than seeking guidance on privacy requirements.

%% file: content/02_relatedwork.tex
\section{Related Work}\label{sec:relatedwork}
This section summarizes related work on the influence of regulations on the software development process, privacy sources used by developers, and the implementation of privacy requirements.

\subsection{Regulations}\label{sec:regulations}
Previous research~\cite{campanile2022towards, marjanov2023data, SaeTheUrb+22, loser2014security,peixoto2023perspective,alhazmi2021m} provided comprehensive insights into how GDPR and similar regulations shape the software development process. Marjanov et al.~\cite{marjanov2023data} emphasized the necessity of both technical and organizational measures to meet legal standards, identifying specific danger points and conditions that lead to data breaches. This data-driven approach offered developers crucial insights into compliance risks and mitigation strategies. 
Saemann et al.~\cite{SaeTheUrb+22} further explored the impact of regulations through the analysis of GDPR fines, highlighting common compliance failures, particularly during data collection. They advocated for the integration of Privacy-by-Design principles early in the development cycle to enhance data protection strategies and ensure alignment with regulatory standards.
Loser et al.~\cite{loser2014security} examined the influence of GDPR on developer priorities within agile environments, noting that security and privacy are often secondary to features that directly enhance user engagement unless mandated by law. 
These studies underline the significant influence of regulatory frameworks on software design and development.

\subsection{Information Sources}\label{sec:sources} 
Past research explored information sources in the field of software security~\cite{acar2016you, fischer2017stack}. 
For example, Acar et al.~\cite{acar2016you} assessed the quality of Stack Overflow code, finding that the code was often unhelpful and insecure. Fisher et al.~\cite{fischer2017stack} found that 15.4\% of 1.3 million analyzed Android applications contained code from Stack Overflow, with 97.9\% of those containing an insecure code snippet.
Similar observations have been made with privacy-relevant code~\cite{tahaei20understanding, tahaei2022understanding, li2021developers}. For instance, Tahaei et al.~\cite{tahaei20understanding} analyzed the questions developers frequently asked on Stack Overflow. They found that software developers ask for advice on privacy topics such as privacy policies, access control, and decisions on the privacy aspects of the systems they are building. In another study, Tahaei et al.~\cite{tahaei2022understanding} focused their research on the advice posted for developers regarding privacy. They found that the advice predominantly emphasizes compliance with regulations and the implementation of confidentiality measures like encryption and data minimization. 
Further, when analyzing developer posts on Reddit, Li et al.~\cite{li2021developers} found that developers rarely discussed privacy issues unless triggered by external factors, such as app store privacy policy updates.

\subsection{Implementing Privacy Requirements}\label{sec:implementingprivacy}
Prior work provided ample evidence that implementing complex privacy regulations is challenging or lacking~\cite{SaeTheUrb+22,balebako2014improving,senarath2019will,senarath2018understanding,boteju2023sok,ma2024programmer,utz2023privacy}. 
In particular, mobile app developers repeatedly ignore privacy regulations and process personal information way beyond what they are legally allowed to~\cite{NguBacMarSto21,NguBacSto22,KocWesAltOlv+22,KocAltJoh23}.
Implementing privacy measures incorrectly can threaten millions of user data, as shown by the study of Martino et al.~\cite{MarRobWeyQua+19} and their follow-up work in~\cite{MarMeeQua+22}. They analyzed how companies realize the \emph{data subject access request} aspect of GDPR and found that due to insufficient checks of the request, it frequently can be used to leak other people's personal information.


Balebako et al.~\cite{balebako2014privacy} examined the privacy and security practices of smartphone app developers, particularly focusing on challenges in implementing effective privacy policies and data encryption. The research highlights a lack of awareness for privacy measures, e.g., privacy policies and data minimization, among developers, with many relying on social networks and search engines for advice. 


Compared to previous studies that focused on the impact of regulations and developers' adherence to privacy laws from a theoretical or observational perspective, we explored developers' behavior with privacy-sensitive tasks.

%% file: content/03_methodology.tex
\section{Methodology}\label{sec:methodology}
Due to the upcoming popularity of AI assistants and the lack of foundational knowledge of the underlying problem, we opted for a lab study to control the study environment fully, i.e., refraining from a second screen or device or contacting other parties with questions without us noticing. Therefore, we conducted a qualitative study with 30 professional software developers in a laboratory setting.
We focused on GDPR, as it governed our German participants, and companies offering services in the EU must comply with it. While relevant privacy concepts also apply to other regulations (e.g., CCPA), the specifics of compliance may vary. Regulatory differences, such as varying data deletion rights (CCPA~\cite{CCPA} vs. GDPR~\cite{gdprart17}) and consent requirements~\cite{gdprccpacomp}, may influence participant decisions and task performance (see Limitations~\ref{chap:limitations} for details).
Participants were divided into the following 3 groups. 


\textbf{Control (C):}
First, we explored if developers would think about privacy-compliant implementation on their own by not mentioning privacy in the task description. 
Participants in the Control Group were presented with a standard task description that was devoid of any explicit mention of privacy considerations. 

\textbf{Prompted (P):}
Second, we investigated if privacy requirements might be too complex for implementation by explicitly asking participants to ensure a privacy-compliant solution. 
This group received explicit instructions to ensure their solutions adhered to privacy regulations and user data protection in the beginning and the end of the task description, provided as a pdf file on the study computer.

\textbf{Expert Support (E):}
Third, we tested if implementation might improve with privacy expert support, addressing the communication gap found in past research~\cite{horstmann2024those}, where developers reported a lack of support on privacy requirements. 
Thus, participants in the Expert-Support Group were prompted for privacy compliance and offered to contact a privacy expert through a chat function. The expert was responsible for providing individual consultations, while ensuring consistent responses to identical questions. Technical issues, such as reversing a participant's accidental deletion of user data, were beyond the expert’s scope.


We analyzed participants' use of privacy sources and their compliance based on the implementation's source code. After completing their tasks, we conducted semi-structured interviews with the participants, analyzing their thoughts on privacy, privacy sources, and privacy issues they had during the implementation. 
While we did check if the solutions produced by the participants were functional, e.g., would run and fulfill the stated purpose of each task, the research goal was to see if privacy was considered during the task.
We summarize the essential parameters of our study in Table~\ref{table:study_parameters}~\cite{serafini2024engaging}.
\input{tables/study_parameters.tex}

\subsection{Study Application Prototype}
Consulting one researcher with professional software development experience and another with data protection consultant experience, we designed a scenario with sensitive health data. This was rated highly relevant to real-world applications, covering a broad range of GDPR principles and requiring additional safeguards.
Thus, we developed a prototype of a health application inspired by existing health apps~\cite{zocdoc,doctolib}. 
The app's primary function was for users to schedule and manage doctor appointments. The application encompassed vital features such as a search function to find doctors by name and specialization, an appointment request system, health record management, a calendar view for appointment scheduling, a dashboard overview, and a functioning backup system. 
Doctors can request access to patients' health records and, if granted, gain access to prior treatments and medication. 
The app already had several active users and doctors whose data was randomly generated by the tool Faker~\cite{faker}.
Our application dealt with sensitive user data, such as health data, which ensured that participants had to make privacy-relevant decisions. Some privacy features (e.g., consent management and automated deletion processes for the backup) were intentionally omitted to leave the implementation of privacy-sensitive tasks to the participants.

\textbf{Backend Technologies:}
The application's backend was developed using the Node.js~\cite{nodejs} runtime environment, which provides a robust and scalable platform for server-side development. As our primary web application framework for Node.js, Express.js~\cite{expressjs} facilitates the creation of robust web applications and APIs with a minimalist and flexible set of features. To optimize the deployment and management of our MySQL database and PHPMyAdmin containers~\cite{PHPMyAdmin}, we used Docker Compose~\cite{docker}, a tool that allows the definition and running of multi-container Docker applications. As an integral part of our email processing mechanism, we used MailCatcher~\cite{mailcatcher}, which intercepts and displays outgoing emails from our application.

\textbf{Frontend Technologies:}
The frontend of our application was developed using the Ionic Framework~\cite{IonicFramework}, a platform that enables the creation of cross-platform mobile and web technologies such as HTML, CSS, and JavaScript. As the foundational framework for our frontend application, Angular~\cite{Angular} was used as it provided a comprehensive and structured framework for building dynamic and interactive web applications.

\subsection{Study Tasks} \label{method:results}

Participants received a task and tool description, a list of software used in the application, and an overview of the app functions and their location to make it easier for the participants to navigate the application.
We included one startup task (\task{task0}) and four study tasks in the study. \task{task2}, \task{task3}, and \task{task4} had privacy compliance issues according to the GDPR. All Tasks had clear finish conditions provided to the participants.
The full text of the tasks can be found in the supplementary material (see Section~\ref{sec:availability}).
\newline

\textbf{\deftask{task0}: Running the Application.} This task guided participants through the process of starting the application. It was intended as a warm-up and to provide participants with a quick way to restart the application in case of a crash.

\textbf{\deftask{task1}: Repair Backup.} This task showed participants what kind of data is stored by the application to ensure they were aware of the existence of the backup function relevant for~\task{task2}. It required them to finish the backup functionality of the application, which did not store one of the tables.
To solve this task successfully, participants had to remove the exclusion clause, which prevented one of the tables from being backed up.
There were no privacy requirements for this task, so we only report the number of functional solutions.

\textbf{\deftask{task2}: Permanently Deactivate Account.} In this task, participants were asked to process a deletion request from one of the application's users.
While removing or anonymizing all personal identifying information from the live database was sufficient to fulfill the functional aspect of the task, \textit{participants were required to delete or anonymize the data in the backup system to preserve the user's privacy correctly.}

\textbf{\deftask{task3}: Search Functionality.} Participants were asked to finish implementing a function allowing doctors to look up past treatments for certain diseases. 
To fulfill the functional requirements, participants had to query the database for patients with the disease in question and display either their medical records or the medications used to treat the disease. While giving health information to a patient's current doctor might not be a violation, the search function without any additional filters would allow all doctors in the system to receive information on every patient.
\textit{A privacy-preserving solution, however, required the participants to either implement some sort of filter such that only the doctor's own patients were displayed or to leverage the existing health record permission system. A solution where no personally identifiable information was shown, e.g., only medications given to treat a disease without the patient's name, was also considered privacy-preserving.}

\textbf{\deftask{task4}: Advertisements.} Finally, participants were required to implement an email function that would send advertisements to the application users.
To solve this task, an automatic email notification system had to be implemented, allowing the administrator to send emails to all application users. 
\textit{A privacy-compliant implementation requires an additional consent system, such as adding an option to the user profile to update their consent or requesting additional permissions from all users.}


\subsection{Interview Guideline}
We designed a semi-structured interview guide consisting of two parts: First, we asked participants whether or not they considered privacy when completing the tasks and their reasons for doing so. If they considered privacy, we asked them how they ensured privacy and what sources they used during the implementation. Participants in the Expert-Support group were explicitly asked for their reasons for contacting or not contacting the expert. Further, participants were asked about the usefulness of their sources.
In the second part of the interview, participants were given the opportunity to reflect on each task, giving their opinion on the privacy issues they felt the tasks posed, how they have or would have ensured compliance, and what support they would like to have for such tasks. Finally, participants were asked if they wished to mention any potential issues or comments not covered during the interview. The interview was conducted in English or German, as requested by the participants.
The programming tasks lasted on average 203.8 minutes (min: 43 max: 303 md: 212 $\sigma$: 72.99), and the interviews lasted on average 12.77 minutes (min: 6.1 max: 29.8 md: 11.83, $\sigma$: 5.08)

\subsection{Piloting}
We further conducted several pilot studies to design a study that is short enough to (i) motivate participation in person, (ii) minimize research costs, and (iii) reflect real-life software development projects that often have to adhere to strict deadlines but long enough to finish the tasks in time.

\paragraph{\textbf{Pilot 1}} We first conducted one pilot study for each group with student assistants from our lab (P\textsubscript{C1}, P\textsubscript{P1}, P\textsubscript{E1}), who had not previously worked on privacy-related projects. P\textsubscript{P1} had professional experience as a web developer. Based on their feedback, we made minor adjustments to the study tasks to make them more comprehensible. Further, as our pilot participants did not solve all the tasks within an initial estimation of three hours, we increased the time limit to five hours and pre-implemented non-privacy parts for \task{task3} to further save programming time. Further, we switched the order of Tasks~\tasks{task3} and \tasks{task4}, since due to the pre-implemented parts in \task{task3}, it was shorter than \task{task4}. Lastly, based on participant feedback, we added \task{task0} to help participants get started with the tool.

\paragraph{\textbf{Pilot 2}} We conducted a second round of piloting with all three groups (P\textsubscript{C2}, P\textsubscript{P2}, P\textsubscript{E2}). 
P\textsubscript{C2} worked full-time as a software developer, P\textsubscript{P2} worked part-time as a security consultant and part-time as a student assistant in our lab, and P\textsubscript{E2}, part-time as a senior software developer. P\textsubscript{P2} was piloting the study as part of their employment in our lab. P\textsubscript{C2} and P\textsubscript{E2} were recruited from a pool of past study participants.   Participants did not manage to solve all the tasks due to time issues and programming difficulties during \task{task4}, "Advertisements." To alleviate the time issues, we formulated clear finish-states in the tasks, avoiding participants spending time on the tasks for unnecessary functionalities (e.g., front-end design of the search functionality). For the programming difficulties, we added an existing email function for doctors and pointed the participants to this function to use as a function to adapt.

\paragraph{\textbf{Pilot 3}} We recruited two more professional software developers from a list of past study participants for the final round of piloting (P\textsubscript{C3}, P\textsubscript{P3}). P\textsubscript{C3} solved the tasks within 3 hours and 15 minutes, and P\textsubscript{P3} within 4 hours and 5 minutes. As the participants could solve the tasks within the required time limit, no further changes to the study were necessary. Thus, we included the participants' results as C1 and P1 in our analysis.

\subsection{Participants}
\input{tables/demographics-aggr}

Recruiting professional software developers for a lab study was challenging due to lack of time, spread-out geographical locations, and high cost. We recruited participants through three different channels: (1) six developers from past computer science studies unrelated to privacy or data protection who agreed to receive future study invitations from our research group, (2) eight through advertising the study through ``Kleinanzeigen``~\cite{kleinanzeigen}, a German Craigslist~\cite{craigslist} equivalent, and (3) 16 by snowball sampling.

We stopped recruitment when no willing participants registered for the study. All 94 interested participants were first presented with a short survey containing the consent form and an outline of the study. Participants were then screened for eligibility using the following criteria: 18 years of age or older, employed in the field of software development, experienced with JavaScript, and passed two out of three screening questions for programmers~\cite{danilova2021you,danilova2022testing}. To prevent participants from answering the questions through Google, we used the time limits for each question as recommended in~\cite{danilova2022testing}. 

Out of 94 participants who started the survey, 74 participants passed the screening criteria. All participants who met our requirements were directed to a calendar application to schedule a study appointment. Thirty of those decided to participate, with some of the remaining ones contacting the researchers to inform them they could not participate due to scheduling conflicts or travel distance. Participants were then randomly assigned to their group when they registered for the study.
We derived the compensation based on past research with software developers, where the average hourly compensation was \num{57}€~\cite{naiakshina2020on}. Applying this rate to our five-hour study resulted in a total compensation of \num{285}€. We added an additional \num{30}€ to cover travel expenses. All participants were compensated with \num{315}€ for their participation in the study. 

The demographics of our participants can be found in Table~\ref{tab:demographics}.
Of the \num{30} participants, \num{24} identified as male and six as female. Participants were 30.8 years old on average (min: 18, md: 27 max: 56, $\sigma$: 10.17). Most (20) had a Bachelor's~(14) or Master's degree~(6). Five achieved a High School degree or equivalent, two had some college experience, two more had a vocational degree, and one had Grad School experience. They worked an average of \num{5.25} years in their current field (min: 0.5, md: 3, max: 25, $\sigma$: 5.85). The majority were employed full-time (18), \num{7} were self-employed, and \num{3} worked part-time. Two participants preferred not to answer the question. The number of employees in the largest company participants had worked in ranged from \num{10} to \num{260000}.


\subsection{Analysis}
In this section, we describe the process of analyzing participants' source code, screen recordings, and interview transcripts. 

\subsubsection{Source Code}

\textit{Functionality:} Two researchers, R1 and R2, independently reviewed the source code the participants created during the study. They checked each task to see if it was functionally fulfilled. 
For \task{task1}, the ``medical\_conditions'' table needed to be saved though the backup function. \task{task2} was considered solved functionally when participants deleted data from the specific user. For \task{task3}, a doctor user needed to be able to enter a medical condition in a search field, which returned some data. Lastly, \task{task4} was considered solved functionally if a mail could be sent through the system and appear in a MailCatcher.

\textit{Privacy:} R1 and R2 went through the source code of the participants, noting if and how the participants considered data protection in their implementations. The findings were reviewed by R3. \task{task1} did not require any considerations for privacy compliance. For \task{task2}, the remaining user data in the database was considered non-compliant. If the backup was changed, they checked if all data connected to the user was gone or anonymized. If it was, the solution was rated as compliant. If only some data was gone, they considered it a partially compliant solution. In \task{task3}, they first checked if personal information was given to the doctor, which might identify the patient (e.g., full names, addresses, emergency contacts). Second, they checked if a filtering method was implemented, ensuring doctors could only see data from their own patients or patients who had given permission to share their data. A solution was for a doctor not to be given access to personal information, but all patients' health data was considered partially compliant. Lastly, for \task{task4}, they considered a solution compliant if a check for consent was implemented.  

\subsubsection{Screen Recording}
The combination of screen recordings and browser history analysis provided insights into both the implementation process and participants' browsing behavior. Each participant's screen was recorded throughout the study, allowing for observation of the task implementation process in addition to the source code analysis. The recordings were initially reviewed at an accelerated pace to identify task implementations and notable events, such as the addition of new features, changes to the user interface, and privacy-related online searches. 
Additionally, the screen recordings enabled a thorough review of participants' browser history, which provided insights into external resources accessed by participants during task execution, particularly those related to privacy guidelines solutions. Any privacy-related searches were noted down. A table was then created to systematically document the findings, including how tasks were implemented, the extent to which privacy aspects were addressed in the implementation, based on the privacy requirements mentioned above, and search activities related to privacy topics in the browser history.


\subsubsection{Interview}
R2 conducted 11 interviews, R4 conducted 19. Transcriptions were done using Amberscript (GDPR-compliant), checked and approved by R2. Researchers R4 and R5 analyzed the interview transcripts using thematic analysis~\cite{braun2006using, clarke2015thematic}. First, the two researchers created a codebook based on two transcripts. The two researchers then coded the interviews individually, 
resolving conflicts through regular meetings and discussions (e.g., missing codes, concept clarification). For example, in these discussions, they added or merged codes introduced by each researcher in the final codebook. After both researchers had coded all the interviews, they resolved all conflicts, for example, differentiating purpose limitation and data minimization, through discussion. Final findings were reviewed with R3. We achieved 100\% inter-coder agreement.
The final codebook can be found in the supplementary material (see Section~\ref{sec:availability}).

\subsection{Positionality Statement}
As suggested in prior work (e.g.,~\cite{Sannon2022PrivacyResearch, Ortloff2023DifferentResearchers}) we provide researchers' backgrounds involved in this work. Researchers specializing in Application Security (R1), Usable Security and Privacy (R2, R4, R5), and Data Protection (R3) conducted data collection and analysis. R3 was additionally a certified data protection officer with relevant work experience and served in expert communication for the expert group.

\subsection{Ethical Considerations}
The study received approval from our university's institutional review board (IRB). Initially, we obtained approval for a 3-hour study with a payment of €200. After piloting, we extended the study duration to 5 hours and increased the payment to €315. These changes were submitted as amendments and subsequently approved by the IRB.
All participants were provided with a consent form outlining the scope of the study, the data collected, the participants' risk, retention policies during the screening process, and a physical copy before the laboratory study. 
We also complied with the GDPR\@, e.g., by encrypting and anonymizing study data and obtaining informed consent from our participants.  
All participants were informed they could stop participating in the study at any time and request the deletion of their data at any point after or during the study.

\subsection{Limitations}
\label{chap:limitations}
Our study has limitations that need to be considered when interpreting the results. 
First, we present the results of a qualitative study. Participants were aware that they were participating in a study and may have behaved differently than in the real world. To address this issue, we used a role-playing technique to counter environmental bias. While we cannot claim generalizability, we provide a first step to understand a complex problem.
Second, all participants were recruited in Germany and thus are not representative of all developers. 
Third, while participants were presented with programming screening questions, not all may have had the required experience level to solve all the tasks. 
Finally, a key limitation of our study is its focus on GDPR. While other legislations incorporate similar concepts (e.g., CCPA), the specifics of compliance may vary. For instance, CCPA allows companies to refuse a deletion request~\cite{CCPA}, whereas GDPR enforces stricter data erasure rights~\cite{gdprart17}. Similarly, health data sharing under HIPAA follows different minimization rules~\cite{HIPAAmin}, and opt-out consent for email ads is valid under CCPA but not GDPR~\cite{gdprccpacomp}. As a result, our findings may not be fully generalized to compliance tasks governed by other regulatory frameworks. Future research is needed to examine how these differences impact participant decision-making and task performance.

%% file: tables/study_parameters.tex
\begin{table}
\caption{Study Parameters.}
\centering
\label{table:study_parameters}
\begin{adjustbox}{max width=\columnwidth}
\begin{tabular}{ll}
\hline
\rowcolor{LightGray} \textbf{Study Type} & Laboratory  \\
\textbf{Study Task}& Programming Task \& Interview  \\
\rowcolor{LightGray} \textbf{Study Language}& English \& German  \\
\textbf{Programming Task Length}& mean: 203.8 min (md: 212, $\sigma$: 72.99)  \\
\rowcolor{LightGray}\textbf{Interview Length}& mean: 12.77 min (md: 11.83, $\sigma$: 5.08)  \\
\textbf{Recruitment Channel}& Past computer science study participants, \\
& Local Job Postings \& Snowball Sampling\\
\rowcolor{LightGray}\textbf{Recruitment Duration}& 4 months  \\
 \textbf{Participants}& Software Developers (n = 30)  \\
\rowcolor{LightGray}\textbf{Compensation}& 315€ per participant  \\
\hline
\end{tabular}
\end{adjustbox}
\end{table}

%% file: tables/demographics-aggr.tex
\begin{table}[t]
    \caption{Demographics of the 30 participants.}
    \label{tab:demographics}
    \centering
    \footnotesize
    \renewcommand{\arraystretch}{0.95}
    \setlength{\tabcolsep}{0.66\tabcolsep}
    \setlength{\defaultaddspace}{0.25\defaultaddspace} 
\begin{adjustbox}{max width=\columnwidth}
    \begin{tabular}{lllll}
        \toprule
       \rowcolor{LightGray} \multicolumn{5}{l}{\textbf{Gender}}  \\
       \rowcolor{LightGray} & Male & 24 (80\%) & Female & 6 (20\%) \\
        \addlinespace
        \multicolumn{5}{l}{\textbf{Age [years]}} \\
        & Min. & 18  & Max. & 56  \\
        & Mean ($\sigma$) & 30.8  ($\pm$10.17) & Median & 27 \\
        \addlinespace
       \rowcolor{LightGray} \multicolumn{5}{l}{\textbf{Experience in current field [years]}}\\ 
       \rowcolor{LightGray} & Min. & 0.5  & Max. & 26  \\
      \rowcolor{LightGray}  & Mean ($\sigma$:) & 5.25 ($\pm$5.85) & Median & 3 \\
        \addlinespace
        \multicolumn{5}{l}{\textbf{Education}}  \\
         & Bachelor's degree & 14 (46.67\%) & Master's Degree & 6 (20\%)\\
         & High School Equivalent & 5  (16.67\%) & Some College & 2  (6.67\%)\\
         & Vocational Degree & 2  (6.67\%) &Some Grad School&1 (3.33\%)\\
         \addlinespace
        \rowcolor{LightGray} \multicolumn{5}{l}{\textbf{Current Employment Status}}  \\
        \rowcolor{LightGray} &Full time &18 (60\%)&Self-Employed   &7 (23.33\%)\\
        \rowcolor{LightGray} &Part time  &3 (10.0\%) &Prefer not to say&2 (6.67\%)\\
         \addlinespace
        \multicolumn{5}{l}{\textbf{Largest Company Size (Employees)}}\\ 
        & Min. & 10  & Max. & 260 000  \\
        & Mean ($\sigma$:) & 19782 ($\pm$59563.79)& Median & 170 \\
         \bottomrule
    \end{tabular}
\end{adjustbox}
\end{table}

%% file: content/04_results.tex
\section{Results}\label{sec:results}

\input{tables/source_code_analysis}
The findings from the code assessment are presented in Table~\ref{tbl:sourcecodeanalysis}.
If the implementation was incomplete but participants at least tried to fulfill privacy requirements based on our criteria outlined in Section~\ref{method:results}, we recognized participants' awareness of privacy compliance as partially fulfilled. All participants fulfilled \task{task0}. 
For all four main tasks, participants could achieve a functional solution, resulting in 40 possible functional solutions for each group. As only three tasks had compliance issues to consider, each group could achieve a maximum of 30 compliant solutions. Special cases where participants did not follow the task as intended were excluded. 
For all tasks combined, while the participants in the different groups delivered a similar number of functional solutions (Control: 28/40, Prompt: 28/38, two special cases, Expert: 30/40), the number of privacy-compliant solutions differed. 
The Control group had the lowest number of compliant solutions (1/30 fully compliant, and 2/30 partially), suggesting that developers could be less likely to consider privacy if they are not explicitly prompted. Next was the prompted group (3/29 fully, 3/29 partly, one special case), with the expert group having the most compliant solutions (7/30 fully, 1/30 partly). While only 4 participants in the expert group contacted the experts, the support of the experts could have potentially supported the participants enough to improve compared to the prompted group. 


\subsection{\task{task1}}
\subsubsection{Source Code Analysis for \task{task1}}
Almost all participants successfully fulfilled \task{task1}. They were capable of identifying and addressing the backup issue, indicating competency in basic application maintenance tasks. Only one participant (P4) failed to complete the task, experiencing potential challenges or misunderstandings in the backup process. Another participant (E10) partially fulfilled the task by identifying the line of code responsible for excluding the "medical\_condition" table from the backup. However, instead of fully addressing the issue, the participant opted to comment out the exclusion code, which led to errors during the backup process. Only one prompted participant did not provide a solution, and another expert participant partially fulfilled it for \task{task1}. 
\begin{summary}{Task 1}
Overall, the task posted little challenges for the participants.
\end{summary}

\subsubsection{In Their Words: Interview Insights for \task{task1}}

Six participants explicitly mentioned they saw no relevant issues with privacy regarding \task{task1}. Four others saw issues with the developers being able to access the data and suggested access control measures or permission systems. Three suggested some form of database obfuscation, such that the data would not be easily readable by developers: \blockquote[P7]{That the names should be not plain readable, the email addresses not. And the other, like more the medicines and so on.}. Two other participants suggested pseudonymization or anonymization to achieve this. Lastly, two participants explained that, in their opinion, not all data should be backed up, making the task itself violating privacy: \blockquote[E3]{Maybe like, yes, not backing up the whole table, but only extract parts from them that are needed for the functionality of the service, but not on the personal data, or ask for permission.}.

\subsection{\task{task2}}
\subsubsection{Source Code Analysis for \task{task2}}
Twenty-five participants functionally fulfilled \task{task2} by successfully deleting all user data from the database. However, only four considered the backups, such as participants E6 and E9, who ensured that user data was also removed from backups, effectively mitigating privacy risks associated with retaining user data.
Twenty-seven participants failed to fully address the privacy aspect of \task{task2}. While they deleted user data from the database, they missed the backups, posing ongoing privacy risks. Further, one participant deleted all existing backups to comply with the data deletion request, inadvertently removing important data backup measures.
While this removes the data of the patient who requested deletion, it is obviously undesirable, as it reverses the benefits of a backup for all other data, potentially leading to data loss.

Participants such as C6 attempted to address backup considerations by selectively deleting user data from specific tables, such as "user\_patient" and "personal\_info," while leaving other tables untouched. While partially addressing backup concerns, this approach raises questions regarding the completeness of data removal and the potential for residual data in other tables.
Similarly, participant C9 deleted user data from only one backup while leaving user data intact in other backups, partially fulfilling the privacy aspect.
Comparably, participant P8 partially fulfilled the task functionally by anonymizing user data in the "user\_patient" table in the backups. However, privacy concerns remained as some backup tables still contained the user's patient ID, indicating partial mitigation of privacy risks despite functional deletion from the database.

Three participants in the control group were unable to provide a functional solution, whereas, in the other two groups, one each was unable to do so. Concerning privacy, in the control group, two participants delivered a partially compliant solution. In the prompted group, one each delivered a compliant and a partially compliant solution. Two participants in the expert group delivered a compliant solution.

Notably, the two participants in the expert group (E6 and E9) successfully fulfilled the task in terms of privacy by ensuring that user data was removed not only from the database but also from backups. Both participants made use of the chat functionality to contact the expert. However, only E6 asked if data needed to be deleted in the backups, with E9 only notifying the expert that they had done so. 
\begin{summary}{Task 2}
While most participants were able to solve the task, privacy was rarely considered, and sensitive user data often remained in the application.
\end{summary}

\subsubsection{In Their Words: Interview Insights for \task{task2}}
Concerning the deletion request in \task{task2}, nineteen participants were aware that data with dependencies must also be deleted. Ten of the participants explicitly mentioned that data in the backup would be relevant and had to be deleted in addition to data in the live database. However, only two did so. Fourteen participants mentioned deleting the data manually. Twelve also suggested an automated deletion process for such cases. Further, six participants mentioned that they, as developers, should not be able to see or edit the application users' data and argued that some form of access control should be in place or data should be obfuscated in some form. Two participants, however, recognized the task as a deletion request, a right under the GDPR. Four participants also mentioned specific considerations on the deletion, such as timely deletion or keeping proof of deletion.
However, two participants believed the user data was important and should not be deleted, which could be a violation if the data is not sufficiently anonymized: \blockquote[C1]{The second task was to delete user data. And from my perspective, this user data shouldn't be deleted because it could be important information for other parties. Not the user itself but for other parties. And yeah, I wrote down that it's not a good idea to delete important user data. It would be better to flag the user as deleted and make it possible for some information that has to be stored to be accessible in the future. But I deleted the user anyway.}. Three participants also discussed anonymizing the users' data so that it could no longer be linked to them as a possible alternative to deletion. 

\subsection{\task{task3}}
\subsubsection{Source Code Analysis for \task{task3}}
Twenty participants successfully fulfilled the functional aspects of \task{task3}, but fifteen of them failed to adequately address privacy concerns, resulting in unfiltered searches across the entire database table. This oversight means that doctors could access confidential patient information without proper authorization. 

Six participants fully integrated privacy considerations into their codebase, while three more partially addressed privacy concerns.
Participants P2 and P3 partially fulfilled the privacy aspect of \task{task3} by refraining from displaying any patient data during searches. However, their implementation allowed doctors to search across all entries in the "medical\_condition" table without considering whether the medical conditions belong solely to their own patients. Although patient identities were safeguarded, doctors could still access comprehensive information about medications and medical conditions without knowing which patient they belonged to. Even if patient identities remain undisclosed, they allow access to potentially identifying data.
Five participants (C2, P5, P6, E3, E4) demonstrated a proactive approach by limiting search options based on patient consent, thereby limiting doctors' unauthorized access to medical records. This approach is consistent with existing application functionality where patients grant permission to access data. They displayed the patient's first and last name and the needed information.

P6 took a privacy-conscious approach by not only limiting search results based on patient consent but also anonymizing the data to increase confidentiality. By excluding patient names from the results and only displaying medical conditions and allergies, the participant prioritized maintaining privacy without compromising functionality.
Similarly, participant E1 limited the search to patients who had previously scheduled an appointment with the respective doctor. This approach improves data protection by ensuring that data access is only granted to patients with previous interactions, thereby minimizing the risk of unauthorized information retrieval.
Six control, seven prompted, and seven expert participants delivered functional solutions. In the control group, one participant provided a compliant solution, compared to two partially compliant and two fully compliant solutions in the prompted group. The participants in the expert group delivered three compliant and two partially compliant solutions. 
\begin{summary}{Task 3}
While functionality was not an issue in this task, most participants did not attempt a privacy-compliant solution or failed to submit a privacy-compliant implementation.
\end{summary}

\subsubsection{In Their Words: Interview Insights for \task{task3}}
For \task{task3}, the implementation of a search function for doctors, twenty-eight participants came to the conclusion that patient data should not simply be available for all doctors. Fifteen participants mentioned some form of access control measures for doctors: \blockquote[C5]{Yeah, I don't think that I should query all the users and be able to see all the medical information from all users. All like, even if I'm the doctor of someone else, I can see the like, I can see a person As if I'm a person Bs doctor, I can see person As private information. That shouldn't be possible, I think.}. Ten suggested minimizing the data the doctor can receive through a search: \blockquote[P5]{The deal here is to treat a certain medical condition and additional data like the first name, last name, birthday, I don't know, blood type is not required in this case. I basically would only give back treatments or medications that have worked positively for this kind of condition in the past. And the rest is not valuable or not relevant for this situation, I would say.}. Six participants saw privacy issues with the tasks but were unsure how or if a compliant solution was possible: \blockquote[C8]{I mean, I don't know, the whole function, like the whole function seems to be a bit, not really privacy-protecting.}. Seven participants discussed consent measures, as participants giving their consent to other doctors might solve the task compliant: \blockquote[C2]{Technically, it's possible. And I'm not sure if it's enough, but at least one has to filter whether for each single user, one has the permissions to read it given by the user, and I would think that the user especially has to allow this kind of use.}.

\subsection{\task{task4}}
\subsubsection{Source Code Analysis for \task{task4}}
Thirteen participants successfully realized the functionality of \task{task4}. Out of those, however, none did so in a privacy-compliant way. 
Several noteworthy implementation aspects that did not quite fit the scoring scheme will be described below. They are denoted as ``Special Case'' (SC) in Table~\ref{tbl:sourcecodeanalysis}.
One participant (P3) misunderstood \task{task4} and added the email functionality to the doctor's account.
Similarly, P8 did implement this task in a way that did not fit the task description either.
Instead of sending it to a list of people, the admin has to type each receiver's email by hand.
Neither E1 nor E9 did finish the task from a functional point of view.
However, the aspects implemented by either highlight that both wanted to check user consent prior to sending e-mails.
E9 added a flag to the user's profile page to opt into e-mail notifications, whereas E1 added a filter for user consent prior to sending e-mails.

Overall, for Task 4, five control participants delivered a functioning solution, and three prompted participants implemented a functioning solution and two special cases. Five expert participants delivered a functioning solution for the last task. No participant in the control group or the prompted group delivered a privacy-compliant solution for the fourth task. In contrast, two participants in the expert group were able to implement a compliant solution. 
\begin{summary}{Task 4}
While the number of functioning solutions was similar for the groups, only the expert group was able to fulfill the privacy requirements.
\end{summary}

\subsubsection{In Their Words: Interview Insights for \task{task4}}
During \task{task4}, twenty-two participants mentioned consent as the main concept to create a compliant solution: \blockquote[E8]{Yes, the parties have all consented to their email addresses being stored there. The question is whether they have also agreed to allow advertising to be sent out.}. Related to this, seven participants mentioned purpose limitation as a concept that was violated by the task: \blockquote[C5]{That shouldn't be, that shouldn't be allowed, I think. Especially that I used email addresses that were for doctor purposes to have for advertisement purposes.}. Similar to \task{task3}, five had concerns with the task itself, seeing it as violating the users' privacy, and were unsure if a compliant solution was possible: \blockquote[P6]{They've probably never even, have never even agreed to receive advertisement emails, so I think the entire task is probably faulty.}. Further, ten participants discussed concepts related to the consent process, such as opt-in/opt-out, informed consent, and technical possibilities to update the consent. Two participants mentioned they did not see an issue with the task, assuming the users would have already given consent when providing their email: \blockquote[P9]{They are allowed to send the email data. So in this case, I think, this should be okay to send the patients emails.}.


\subsection{Privacy Expert Chat}\label{sec:privacy-expert-chat}
Only four (E1, E6, E7, E9) of the ten participants from the Expert-Support Group (E) contacted the privacy expert. Participants asked questions about how to solve the task or whether their idea for a solution was privacy-compliant.

One participant asked if the backup should be encrypted during \task{task1}. For \task{task2}, two participants asked the expert to clarify if data from the backup also needed to be deleted. For \task{task3}, two participants asked about data minimization, purpose limitation, and access control measures for doctors. One participant contacted the expert with a technical question, as they deleted more data from the backup by accident and were unsure how to resolve the issue. Lastly, concerning \task{task4}, two participants asked questions regarding consent. One participant asked the expert to clarify if the users already had consented to advertisements and to confirm that, without consent, the sending of the emails would not be compliant: \blockquote[E1 Chat]{But if a user has not given any consent then sending mails to them would not be privacy-compliant or is it?}. The other participants requested advice on how to best implement a feature allowing the users to give consent, discussing options regarding opt-in and opt-out, as well as the best way for existing users to provide consent: \blockquote[E9 Chat]{Would you consider it to be okay just to send all the existing users the advertisement email, but include a link to opt-out on future emails of that kind? I am personally still concerned about such an approach, but you are the expert regarding privacy topics - so I'm happy to hear your advice}. 
Participants mainly used the chat function to confirm if their own proposed solution was privacy-compliant, as the expert was seen as a trusted source able to evaluate their ideas: \blockquote[E6]{I asked the one time, if all the data, if the patient's data should also be deleted from the backups, but it was the expert, of course, said yes, and it was kind of obvious that it should be deleted}.
One participant who used the chat function the most apologized to the expert for contacting them with questions: \blockquote[E9 Chat]{Thanks again and sorry for bugging you so much lately}. 
While participants who interacted with the expert chat had a positive view of the support, they noted that they wished for clear instructions: \blockquote[E9]{It would have been easier for me if either the task was basically in line with data protection or if I had received more precise instructions from the expert}. 

Three participants, E4, E7, and E10, said they did not need help. However, only E4 delivered a privacy-compliant solution. Two did not contact the expert \blockquote[E2]{Cause I'm kind of afraid of asking too much}, and were unsure what to ask the expert.
Participant E4 noted their priority was the functionality: \blockquote[E4]{That's funny. Yeah. Maybe I was just too lazy and wanted to get the tasks done}. One participant claimed to have missed the prompt for privacy and the chat function with the expert. However, in the screen recordings, the participant was seen reading the prompt and interacting with the chat window initially and after the reminder was sent, indicating an easy excuse for why they did not realize privacy aspects in the allotted time. 
When asked about why they did not seek expert advice on privacy issues, four participants stated during the interview that, in retrospect, they would have contacted the expert more often. \blockquote[E2]{I don't know because I have no... Yeah, I should have used it. Yeah, I think I should have used it.}.


\subsection{Confidence}
We asked participants to rate their confidence that the solutions they provided would be compliant with data protection regulations. 
One participant was extremely confident, and two were very confident. Eight were moderately confident, seven slightly, and 12 were not confident at all. On average, the participants in the control group were the least confident in their solutions, being on average between ``Not confident at all'' and ``Slightly confident'' (min: 1, mean: 1.5, md: 1, max: 3, $\sigma$: 0.85). The prompted participants were a little bit more confident and were slightly confident on average (min: 1, mean: 2.3, md: 2, max: 5, $\sigma$: 1.16). Lastly, the Experts were only a little bit more confident compared to the prompted group, being, on average, exactly between slightly and moderately confident (min: 1, mean: 2.5, md: 3, max: 4, $\sigma$: 1.18).  

The number of compliant solutions differed based on confidence; however, it did not scale up linearly according to the confidence of the participants. We evaluated partial compliant solutions as 0.5 compliant solutions and fully as 1.
The extremely confident participants (P1) did not deliver any compliant solution. The slightly confident participants (C7, P4, P6, P7, P8, P10, E8) performed a bit better; however, they still had on average (0.21) less compliant solutions than the not at all confident group (C1, C2, C3, C4, C5, C6, C8, P3, P5, E3, E5, E7) had (0.42). The moderately confident participants (C9, C10, P2, P9, E1, E2, E4, E6) had, on average (0.75), the second most compliant solutions. The very confident participants (E9, E10) had, on average (1.25), the most compliant solutions. 
\input{tables/participants}

Twenty participants found the programming tasks moderately challenging. Two (P6, E5) found them not challenging at all, and three (C1, C3, E3) found them slightly challenging. Four (C7, P5, E1, E6) found the tasks very challenging, and for one participant (C4), the tasks were extremely challenging. 
An overview of the self-evaluation for each participant can be found in Table~\ref{tab:participants}.

\subsection{Screen Recording Analysis}
Despite the emphasis on privacy compliance tasks, only a few participants relied on Google to find information regarding privacy and actively sought external resources related to privacy regulations. Notably, the majority of online searches were related to technical questions rather than privacy-related concerns. 

Only four participants looked for additional information on privacy. P2 broadly looked for information on consent with the search term ``GDPR consent,'' indicating an attempt to seek information regarding GDPR consent requirements. However, no further exploration or implementation of privacy-related functionalities was observed. C4 demonstrated a more proactive approach to addressing privacy concerns and looked up information with the terms ``medical records GDPR,'' and ``health data GDPR.'' 
Within the ``people also ask'' section of Google, they read the answer to ``Is health data personal data?'' and lastly, they opened Recital 54 (Processing of Sensitive Data in Public Health Sector) of the GDPR. 

E10 looked up information on ``how to make sure every specific user data is deleted in a database?'' and accessed one post each on the topic on Quora.com and auth0.com to gather insights on data deletion practices and strategies employed by developers. Lastly, P8 searched ``How to delete data privacy compliant'' and opened a web source from dataprivacymanager.net, which provided comprehensive guidance on data deletion procedures in compliance with GDPR regulations. Interestingly, P8 was the only participant using a Large-Language model. P8 asked BING Copilot the same question to get further clarification on GDPR compliance. They received an answer detailing Data Retention and GDPR, Operationalizing Data Removal, Complex IT Environments, Right to be Forgotten, and Practical Steps for GPDR-Compliant Data Removal. The full response by BING Copilot is provided in the supplementary material (see Section~\ref{sec:availability}). 

\subsection{Participants' Explanation for Privacy Behavior}\label{sec:interview-explanation}
Participants mentioned different problems during the implementation, which made it difficult for them to ensure privacy compliance. 

\textbf{Relying on Existing Implementation.} Ten mentioned existing implementations within the app, which they deemed not to be their responsibility to address: \blockquote[E2]{I just found something, but I haven't written it because it's already in the project, so it's not my fault}, indicating that participants while noticing privacy issues, decided not to address them as they were not seen as part of a developer's responsibility. For example, one participant commented on \task{task4}: \blockquote[E5]{I'm not sure about whether it's okay to send everyone an email just randomly. So I don't know whether you had to make some subscription to something like a newsletter or something because, yeah, I don't know. And I think this application doesn't have something like that. So I was straightforward and okay, you want it, and then you get it.}. Similar to this, eight participants looked at existing features in the implementation. They used the existing privacy consideration as guidelines, trusting the past implementation would be compliant: \blockquote[E7]{I mostly wanted to just work with the scheme and the functionality that I had. And the application already had a permission system running and working, so I just kept my solutions to that and hoped that this would already be compliant.}.

\textbf{Lack of Knowledge.} Others mentioned their lack of knowledge influencing their privacy decision. Fourteen participants mentioned issues with inexperience on technical issues during the implementation: \blockquote[C7]{Yeah, I did google a lot of stuff to understand how to work with Angular and implement the features. And that might be problematic to privacy compliance because I'm looking. First of all, I don't know the framework really well. So I'm more inclined to copy solutions, copy code, copy snippets and then not know if that will be a risk.}. 

\textbf{Lack of Time.} For four participants, this resulted in a lack of time, causing them to choose a functionality-first approach in their solution: \blockquote[E6]{In the first few tasks, I was really confident that it's data secure. And then the last ones, I had problems implementing the things. So, I think I would have started looking for this if I finished the tasks. Yeah, if I got the functionality right.}.

\textbf{Lack of Experience.} Six others specifically mentioned a lack of experience with privacy and data protection as an issue: \blockquote[P8]{Because if I'm being completely honest, I don't know exactly how I should implement this properly in terms of data protection, something like that or what is okay or not okay.}.
Others ignored privacy for various reasons. Five claimed tasks themselves violated the users' privacy, making it difficult for them to solve the tasks compliantly: \blockquote[P8]{That was a really big no-no, I thought. I implemented the function anyway, or at least an attempt at it, and yes, I will say, I find the idea very difficult anyway, but if you really want to have it in, then you would actually have to implement it in such a way that... No, I don't know how to do this in a data protection-compliant way. In my opinion, the basic idea is somehow gone.}. 

\textbf{Lack of Expert Assistance.}
Three participants (E4, E5, E8) mentioned that they do not have access to privacy experts at work. Participants also talked about problems they had when communicating with experts. For example, E7 mentioned they would prefer the expert to have technical knowledge: \blockquote[E7]{But I think it would be in general better if the lawyers knew concepts of IT and new base functionalities of databases and other things like network traffic. So they have a better idea.}. Two participants mentioned communication with experts at work would be difficult and that they lacked proper procedures: \blockquote[E8]{Also, the communication doesn't really come up since we don't really follow all the proper [procedures].} One mentioned many of their colleagues would not be aware that there are experts they can contact with questions: \blockquote[E10]{So I think it could certainly be promoted a little more, in the sense of: ``Hey, there's someone you could write to.'' Well, I've had to deal with him now and then, but I don't think that all of my colleagues have that kind of contact.}. {Two participants talked about their worries about legal repercussions. They mentioned that they would prefer to refuse to implement certain tasks in real life: \blockquote[C4]{I would prefer to delegate such tasks directly to other people. I don't feel like it, I'll be honest. My head is just way too invested in it, and I don't want to be prosecuted if there is any small mistake. Then you're done.}. Another mentioned difficulties because the lawyer they could contact did not have an IT background, and another noted that developers lacked formal privacy training. Thus, four participants explicitly mentioned that they would like to see more training and education, especially on privacy.

Ten participants claimed they would likely behave the same way in real life as in the study. \blockquote[E6]{Maybe, I think I would have been more hesitant to delete the database, but that doesn't... I don't think so.}. 
Five participants from all three groups explicitly mentioned they wished for the possibility to contact experts with privacy questions. Six (e.g., C4, P6, E3, E10) claimed they would contact an expert or a legal team with questions regarding privacy in real life. However, of the ten participants who had access to an expert, only four made use of this option. Interestingly, both E3 and E10 did not use the expert during the study. E3 reported implementing a functional solution first but that, in hindsight, they should have contacted the expert. E10 noted in the interview that they did not need help with the tasks they solved but would have contacted them during the other tasks. Further, seven participants mentioned they would raise concerns regarding privacy or security issues to superiors or colleagues: \blockquote[C2]{And I wouldn't even be that surprised if the CTO would have requested to implement a feature like task four so the advertisement emails. I would definitely have said that this is not conforming to the regulation and that this feature can't be implemented in that way.}. One participant claimed they would expect to have received training on privacy if they had to implement such tasks. \blockquote[C7]{When it comes to my real life work, if I were to work in an organization, I would expect to be trained or, like, taught exactly what's sensitive, what's not sensitive}. Further, three participants noted it would be good to have measures in place that could prevent privacy violations, e.g., through access permissions given to the employees working with the data: \blockquote[C7]{And obviously, the organization itself would put restraints on my account on my profile}, or preventing them from using certain software. 

\textbf{High Effort.} One participant perceived a compliant solution as too much work: \blockquote[C5]{Because I was too lazy in most of the functions, especially those get all users were already implemented, I just used them straight out of the box. I was thinking about that it might not be the best idea, but I was too lazy at that point.}.

\textbf{Lack of Privacy Advice.}
While participants rarely used online resources during the study, in the interviews, they mentioned sources they used in real life and their educational experiences with privacy issues.
Ten participants mentioned they generally had not used any sources to answer privacy questions.
Four mentioned that they would consult GDPR for privacy issues, and two mentioned rules at work that are used as guidelines. Customers, policies, Stack Overflow, Google, and their education were mentioned by one participant each. One participant specifically mentioned not being sure where to look for privacy questions: \blockquote[C1]{But for how to implement such a feature again, like Stack Overflow or Google, for the privacy part, I don't know}. C1 did not deliver privacy-compliant solutions.

Participants mentioned specific challenges they had when using sources for privacy and privacy requirements in the past. Generally, two mentioned issues with sources written in legal language. Participant C4, who looked for information on health data and accessed ``Recital 54: Processing of Sensitive Data in Public Health Sector'' of the GDPR had the following to say: \blockquote[C4]{Like most things, it is very much legalese. If you don't have the time to find parallel sources, then it's really difficult to understand what the state actually wants and what it wants to express with it.}. Participants explained they found sources confusing and would need a lot of time or consult additional sources to know what is required of them. \blockquote[P8]{ [...] it was a little bit my impression that it would take quite a long time if I worked my way through it now, and that's maybe what I have right now. There wasn't enough time left, and that's why I left it alone. Well, I'll say, at least at first glance, it's not really clear what you have to stick to if I'm being honest.}.

Thus, participants explained they wished for clearer guidelines containing clear dos and don'ts and covering common privacy issues: \blockquote[E10]{Yes, well, maybe a guideline like that, or especially for such, more common data protection use cases, I would say, where you can perhaps look it up briefly, that might be something.}. 
In general, they wished for guidance on privacy issues during implementation; however, they stated that they would ideally have clear to-do's and examples in the guidance: \blockquote[C2]{What would be, in my opinion, the most helpful would be, for example, for the backup system, some official page or even semi-official page where an example backup system is provided for an abstract technical implementation and the steps that one has to take to make sure that the data is saved in a law-abiding way.}.

\textbf{Security vs. Privacy.}
Twenty-two participants mentioned security considerations instead of thinking about privacy issues. 
Sixteen of the participants commented on encryption: \blockquote[P5]{Of course, they need it to be backuped, but I would probably encrypt the backup so that the data is safe, of course, but it can be accessed from like, easily accessed from somewhere.}. Similarly, seven participants were often more concerned with potential attackers or security issues instead of privacy: \blockquote[P7]{I'm more on the other side to make the application safe for hackers and something like that, not data privacy}. Other concerns participants raised were SQL injections and insecure endpoints. They also mentioned concepts such as penetration testing or static code analysis.
Notably, participants noticed some privacy violations during the implementation, enabling them to raise concerns to their superiors. 
Developers who valued user privacy often experienced frustration when required to implement tasks they perceived as privacy violations. This frustration led participants to express their discontent using strong language, such as:
\blockquote[P2]{I began to implement the function about sending the email advertisements, and I stopped mid-implementation and to just remember what I, what the f**k I'm going to do. This isn't okay. So it takes some time to get accustomed to such situations.}.


%% file: tables/source_code_analysis.tex
\begin{table}
	\begin{threeparttable}
 \caption{Source Code Analysis.}
 \label{tbl:sourcecodeanalysis}
 \footnotesize%
 \setlength\tabcolsep{3.5pt} 

\begin{tabularx}{\columnwidth}{l|l|l|l|l|l|l|l|l}
 &  \multicolumn{1}{c|}
{\task{task1}} & \multicolumn{2}{c|}{\task{task2}}      & \multicolumn{2}{c|}{\task{task3}}      & \multicolumn{2}{c|}{\task{task4}}&Time\\
            & Func. &  Func. & Privacy & Func. & Privacy & Func. & Privacy &h:m  \\
            \cmidrule{1-9}%
           \rowcolor{LightGray} {C1} & \cmark{}  & \cmark{}            & \xmark{}                   &    \cmark{}        &  \xmark{}                  & \cmark{}            & \xmark{}        &    3:15     \\
            {C2} & \cmark{}  & \cmark{}  & \xmark{} & \cmark{}  & \cmark{}  & \cmark{}  & \xmark{}& 2:35\\
           \rowcolor{LightGray} {C3} & \cmark{} & \cmark{} & \xmark{} & \cmark{} & \xmark{} & \cmark{} & \xmark{} & 3:30\\
           {C4} & \cmark{} & U & \xmark{} & U & U & U & U & 1:45\\
          \rowcolor{LightGray}  {C5} & \cmark{} & \cmark{} & \xmark{} & U & \xmark{} & \cmark{} & \xmark{} & 4:35\\
        {C6} & \cmark{} & \cmark{} & P & \cmark{} & \xmark{} & \cmark{} & \xmark{} & 3:52\\
          \rowcolor{LightGray} {C7} & \cmark{} & \cmark{} & \xmark{} & \cmark{} & \xmark{} & U & U & 4:28\\
          {C8} & \cmark{} & \cmark{} & \xmark{} & \cmark{} & \xmark{} & U & \xmark{} &4:41 \\
         \rowcolor{LightGray}  {C9} & \cmark{} & \xmark{} & P & U & U & U & U & 1:11\\
        {C10} &  \cmark{} & \xmark{} & \xmark{} & U & U & U & U &0:55 \\
        \rowcolor{LightGray}    Sum & 10 & 7 & 1 & 6 & 1 & 5 & 0 & 30:07 \\
            \cmidrule{1-9}
        \rowcolor{LightGray}     {P1} & \cmark{} & \cmark{} & \xmark{} & \cmark{} & \xmark{} & \xmark{} & \xmark{} &4:05 \\
           {P2} & \cmark{} & \cmark{} & \cmark{} & \cmark{} & P & U & U &3:37 \\
        \rowcolor{LightGray}    {P3} & \cmark{} & \cmark{} & \xmark{} & \cmark{} & P & SC  & \xmark{}  & 3:11\\
           {P4} & \xmark{} & \cmark{} & \xmark{} & \cmark{} & \xmark{} & \cmark{} & \xmark{} & 4:08\\
        \rowcolor{LightGray}    {P5} & \cmark{} & \cmark{} & \xmark{} & U & \cmark{} & U & U &3:09 \\
            {P6} & \cmark{} & \cmark{} & \xmark{} & \cmark{} & \cmark{} & \cmark{} & \xmark{} & 2:50\\
          \rowcolor{LightGray}  {P7} & \cmark{} & \cmark{} & \xmark{} & \cmark{} & \xmark{} & \cmark{} & \xmark{} & 4:17\\
            {P8} &  \cmark{} & \cmark{}  & P & \xmark{} & \xmark{} & SC & SC & 4:35\\
        \rowcolor{LightGray}    {P9} &  \cmark{} & \cmark{}  & \xmark{} & \cmark{} & \xmark{} & U & U &2:45 \\
            {P10} &  \cmark{} & U & U & U & U & U & U & 0:43\\
      \rowcolor{LightGray}       Sum & 9 & 9 & 1.5 & 7 & 3 & 3 & 0 & 33:20 \\
            \cmidrule{1-9}
      \rowcolor{LightGray}  {E1} &  \cmark{} & \cmark{} & \xmark{} & \cmark{} & \cmark{} & U & \cmark{} & 4:57\\
       {E2} & \cmark{} & \xmark{} & \xmark{} & \xmark{} & \xmark{} & U & \xmark{} & 3:56\\
    \rowcolor{LightGray}    {E3}  & \cmark{} & \cmark{} & \xmark{} & \cmark{} & \cmark{} & \cmark{} & \xmark{} & 3:28\\
      {E4}  & \cmark{} & \cmark{} & \xmark{} & \cmark{} & \cmark{} & \cmark{} & \xmark{} & 2:26\\
     \rowcolor{LightGray}   {E5} & \cmark{} & \cmark{} & \xmark{} & \cmark{} & \xmark{} & \cmark{} & \xmark{} & 1:56\\
    {E6} &  \cmark{} & \cmark{} & \cmark{} & \cmark{} & \xmark{} & \cmark{} & \xmark{}&  5:03\\
      \rowcolor{LightGray}  {E7} & \cmark{} & \cmark{} & \xmark{} & U & \xmark{} & U & U &3:34\\
      {E8} & \cmark{} & \cmark{} & \xmark{} & \cmark{} & \xmark{} & \cmark{} & \xmark{} & 2:53\\
       \rowcolor{LightGray} {E9} & \cmark{} & \cmark & \cmark{} & \cmark{} & P & U & \cmark{} &5:01 \\
   {E10} & P & \cmark{} & \xmark{} & U & U & U & U&   4:33 \\
       \rowcolor{LightGray}  Sum & 9.5 & 9 & 2 & 7 & 3.5 & 5 & 2 & 37:47 \\
            \cmidrule{1-9}%
\end{tabularx}
\begin{tablenotes}
\item \cmark: Fulfilled, \xmark: Not fulfilled, P: Partially fulfilled, U: Did not complete, SC: Special Case, did no follow task as intended
\end{tablenotes}

\end{threeparttable}
\end{table}

%% file: tables/participants.tex
\columnbreak


\begin{table}[htbp]
\centering
\caption{Overview of the 30 participants.}
\label{tab:participants}
\footnotesize

\setlength{\tabcolsep}{1pt}
\begin{adjustbox}{max width=\textwidth}    
\begin{tabular}{llccccc}
\textbf{} & \textbf{Current Job Title} & \makecell{\textbf{JS}\\\textbf{Skill}\\}  &\makecell{\textbf{Data}\\\textbf{Privacy}\\\textbf{Skill}}&\makecell{\textbf{Company}\\\textbf{Employees}}&\makecell{\textbf{Task}\\\textbf{Difficulty}}&\makecell{\textbf{Privacy}\\\textbf{Trust}}\\ 
\midrule
{C1}          &Software Dev 
&4
&3&1-99&2&1
 \\
\rowcolor{LightGray}{C2}          & Software Engineer
&4
&3
&1-99
&3&1
\\
{C3}          &Application Dev

&3
&3
&1-99
&2&1
\\
\rowcolor{LightGray}{C4}          &Student Trainee
 
&2
&1
&1-99
&5&1
\\
{C5}          &Software Dev
 
&3
&1
&1000+
&3&1
\\
\rowcolor{LightGray}{C6}          &IT-Consultant
 
&4
&2
&100-999
&3&1
\\
{C7}          &Full-Stack Dev
 
&3
&2
&1-99
&4&2
\\
\rowcolor{LightGray}{C8}          &Working Student

&2
&1
&1000+
&3&1
\\
{C9}          &Apprenticeship
 
&4
&3
&100-999
&3&3
\\
\rowcolor{LightGray}{C10}         &Student
 
&2
&2
&100-999
&3&3
\\\midrule
{P1}          & Software Dev Java

&2
&4
&1000+
&3&5
\\
\rowcolor{LightGray}{P2}          &DevOps Architect

&3
&4
&1000+
&3&3
\\
{P3}          &Software Dev
 
&2
&2
&1-99
&3&1

\\
\rowcolor{LightGray}{P4}          &Software Dev

&2
&2
&1000+
&3&2
\\
{P5}          &Software Dev Frontend

&3
&2
&1000+
&4&1
\\
\rowcolor{LightGray}{P6}          & Assistant Dev

&3
&2
&1-99
&1&2
\\
{P7}          &Senior Software Dev
 
&4
&3
&100-999
&3&2
\\
\rowcolor{LightGray}{P8}          &Software Dev
 
&3
&1
&1000+
&3&2
\\
{P9}          &PHP \& JavaScript Dev
 
&4
&3
&1-99
&3&3
\\
\rowcolor{LightGray}{P10}         & Student

&2
&1
&1-99
&3&2
\\\midrule
{E1}          & Software Engineer

&4
&3
&1000+
&4&3
\\
\rowcolor{LightGray}{E2}          &Junior PHP Dev
 
&4
&3
&1-99
&3&3
\\
{E3}          &Cyber Security Engineer
 
&3
&4
&100-999
&2&1
\\
\rowcolor{LightGray}{E4}          &Senior Consultant 
 
&3
&3
&1-99
&3&3
\\
{E5}          &Software Engineer
 
&5
&3
&1-99
&1&1
\\
\rowcolor{LightGray}{E6}          &Student
&3
&1
&100-999
&4&3
\\
{E7}          &PHP-Developer
 
&2
&2
&100-999
&3&1
\\
\rowcolor{LightGray}{E8}          & Application Dev
 
&3
&2
&1-99
&3&2
\\
{E9}          &Software Engineer
 
&4
&4
&1-99
&3&4
\\
\rowcolor{LightGray}{E10}         &Web Dev
 
&3
&3
&1000+
&3&4
\\
\bottomrule
\end{tabular}
\end{adjustbox}
\\
\footnotesize{C = Control, P = Prompt, E = Expert Chat + Prompted}\\
\footnotesize{Not at all (1) -- Extremely (5)}
\end{table}

%% file: content/05_discussion.tex
\section{Discussion}\label{sec:discussion}
In this section, we discuss our results based on our research questions. 

\textbf{RQ1: Privacy Implementation.}
The overall results of compliant solutions were alarmingly low.
While the number of functional solutions indicated similarity within the three groups for all four tasks (Control: 28/40, Prompt: 28/38, Expert: 30/40), the number of compliant solutions was higher for the prompted group and the expert group. 
For the three tasks with privacy-compliance issues, with partial compliant solutions counted as 0.5, the control group delivered the least compliant solutions overall (2/30), the prompted group delivered slightly more compliant solutions (4.5/29), and the experts the most (7.5/30).

During tasks involving data handling, such as the deletion request in Task 2, participants generally recognized privacy concerns but did not consistently implement robust solutions. This was particularly evident in their handling of backups, where most participants failed to adequately address privacy, leaving significant gaps in data protection.
For tasks that required nuanced data management, such as enabling search functionality for patient data (Task 3), many participants implemented broad access controls without strong privacy protections. This often resulted in solutions that acknowledged privacy but did not adequately protect it.
Overall, our study revealed a mismatch between understanding privacy requirements and putting them into practice, particularly evident in Task 4, where despite attempts to integrate consent checks before sending emails, the overall execution often fell short of privacy compliance. 

These challenges underscore a broader pattern where functional requirements overshadowed comprehensive privacy considerations.
Some developers even clearly stated they would like to avoid privacy-related tasks, as they perceived the subject as challenging and felt mistakes on their side might lead to repercussions as far as being sued. 
While it seems prompting the participants and giving them access to expert support could improve the number of compliant solutions, their number was still very low across all three groups. 

\textbf{RQ2: Usage of Privacy Advice.}
The availability of a privacy expert did not significantly alter the results. Despite access to expert advice, engagement was minimal, suggesting possible communication barriers or perhaps a cultural norm within organizations that may not prioritize privacy consultations as part of the standard development process.

While participants from all three groups mentioned they would like to have support from experts, giving ten participants the option to contact an expert during the study revealed that participants rarely use such options, with less than half of them using the option at all. Mostly, participants only confirmed their own solutions to be privacy-compliant. Further, participants who made the most use of this option in our study even apologized for ``bothering'' the expert. 

The field of psychology recognizes expert-layperson communication as challenging~\cite{bromme2000beyond,bromme2005barriers}, and similar issues have been observed in the field of security~\cite{poller2017can,thomas2018security}.
For instance, scientific literature often argues for more security advice and more cooperation with security experts (e.g.,~\cite{gutfleisch2022does,weir2020from,busse2019replication}). However, our study suggested that access to privacy experts has only a small effect on privacy tasks. Several reasons might influence the participants' behavior in this regard. First, past studies suggested that developers often expressed feeling restricted by privacy requirements in their implementation~\cite{horstmann2024those}. Second, privacy experts have been described by our participants as lacking technical knowledge of implementations. This also extends to the privacy requirements participants got from these experts, which have often been described as hard to understand and written in legal language, lacking clear instructions for the developers (see Section~\ref{sec:privacy-expert-chat}).

Despite the emphasis on privacy-compliance tasks, only a minority of participants actively sought information about privacy regulations. Instead, most online searches focused on technical questions concerning functionality, suggesting a possible gap in understanding or prioritizing privacy issues during development.
Further, some participants mentioned issues they had experienced with sources. Sources written in legal language were deemed too time-consuming to understand and often not specific enough for developers to apply to their tasks directly.

\textbf{RQ3: Privacy Implementation Issues.}
Although participants recognized issues during programming, they often did not address privacy violations for various reasons. 
Some participants mentioned that the software's existing privacy methods, or the lack thereof, influenced their privacy considerations. 
In some cases, participants specifically indicated that they would adopt the same level of privacy as the existing implementation.
Participants also mentioned that they assumed privacy issues had already been addressed, for example, by obtaining user consent. 
Others did not see privacy as their responsibility, choosing to proceed with the task despite being aware of privacy concerns. 
In particular, participants referred to a lack of knowledge, time, expert assistance, privacy advice, and high effort as issues they encountered with privacy-compliant implementation (see Section~\ref{sec:interview-explanation}).

In fact, \task{task3} resulted in the most privacy-compliant solutions, i.e., \num{6}/30 fully and \num{3}/30 partially compliant solutions.
This might be caused by the option to directly reuse existing functionality for consenting to provide access for doctors to a patient's health record, which was used by \num{5}/30 participants. 
Both \task{task2} and \task{task4} required more effort by participants.
For the former, they would have to edit existing backups. For the latter, they would have needed to implement a new consent functionality from scratch.
This mirrors participants' interview responses (cf. Section~\ref{sec:interview-explanation}), where some preferred to reuse existing functionality.
Further, they noted that implementing a privacy codebase might be difficult at this stage, claiming it should have been implemented earlier and that they did not consider it part of their tasks, referring to Privacy by Design solutions.


Overall, as privacy is not seen as a functional requirement, motivation to implement privacy is low, being treated as a ``hygiene factor,'' even for developers who see privacy as important. In large companies, developers' low motivation for privacy may be mitigated through organizational measures (e.g., audits, high-level implementation). However, software is often built by small teams or individuals, where privacy decisions fall to developers. Our programming task revealed that developers struggle to understand and implement privacy requirements, even when prompted. Follow-up interviews identified key issues: (1) privacy ranked below functionality and security, even when violations were recognized, (2) privacy seen as restrictive, (3) difficulty handling privacy sources, and (4) reluctance to consult privacy experts. To address these issues, future work could examine the role of software architects and project managers in ensuring privacy-compliant software development. Potential solutions could include clear guidelines from architects, early privacy education, and project managers encouraging developer-expert communication. Additionally, improving privacy sources and expert guidance—e.g., avoiding legal jargon—could make privacy more accessible for developers.

%% file: content/06_conclusion.tex
\section{Conclusion}\label{sec:conclusion}

With this work, we empirically explored why software developers struggle to implement privacy requirements by conducting a qualitative laboratory programming study with 30 software developers, split into control, privacy-prompted, and expert-supported groups. We asked participants to complete four programming tasks for a health application. We found participants from all three groups struggled to develop their software in accordance with privacy compliance.
Our key insights are:

\textbf{Supporting Developers.} Explicitly prompting developers to privacy and supporting them with privacy experts only slightly improved the compliance of the solutions. However, developers still hesitate to contact privacy experts. Future research would need to explore whether rapport with the privacy expert might influence the behavior of contacting the expert.

\textbf{Functionality First, Security Second, and Privacy Third.} 
Many participants showed a functionality-first approach referring to security when asked about privacy issues, with a tendency to adapt a (non)existing level of privacy implementation. However, focusing on potential attackers often does not help user data issues. We argue privacy and security should not be treated as the same issue, as both usually require independent considerations and measures. We propose conducting more research on how to improve privacy-compliant implementation.
While Privacy by Design~\cite{cavoukian2009privacy} was proposed to be a guiding framework, more research is needed with tools suggesting to support developers with privacy-compliant implementation~\cite{AndMahWhiEnc+20, SheSuKanPha+23,li2018coconut, li2021honeysuckle, FerBriTia+23, KleRolBarKar+23, SrdBasHub23}.

\textbf{Information Sources.}
Participants were uncertain about where to look for information and noted that the legal text they found online was unsuitable for developers. While most of our participants did not use AI assistants, we recognize their upcoming popularity. More research is needed to determine whether these tools might be perceived as helpful with privacy-compliant implementation.

%% file: content/07_availability.tex
\section{Availability}\label{sec:availability}
To ensure transparency and reproducibility, we provide the following items: (i)~source code of the \emph{health-app} with all study materials provided to the participants, as well as a replication package containing the (ii) task description, (iii)~Bing Chat response on privacy, and (iv)~codebook of our qualitative analysis under: \url{https://github.com/privacy-programming-study/project-health-app}. 

The (v)~interview guide~\ref{apx:interview}, (vi)~screening survey~\ref{apx:screening}, and (vii)~post-survey~\ref{apx:post} 
are provided in the Appendix. 
To protect our participants' privacy, we do not provide the interview transcripts or source code solutions handed in by the participants.

%% file: content/08_acknowledgement.tex
\section*{Acknowledgments}
Funded by the Deutsche Forschungsgemeinschaft (DFG, German Research Foundation) under Germany's Excellence Strategy - EXC 2092 CASA - 390781972.

%% file: content/a5_interview.tex
\section{Interview Guide}\label{apx:interview}

Can you briefly tell me which tasks you solved?
\newline
If Control Group: 
\begin{itemize}
\item The project you were working on concerned sensitive user data. Did you consider data privacy compliance in your solution?  Why/why not?
\item If Privacy was considered:
\begin{itemize}
\item How did you ensure data privacy compliance?
\begin{itemize}
\item Did you follow any guidelines or online sources?
\end{itemize}
\item What challenges did you face during the implementation w.r.t. privacy?
\item Did you use any online sources to help you with your challenges w.r.t. Privacy?
\begin{itemize}
\item What sources did you use?
\item How would you rate their effectiveness?
\item How do you feel they could be improved?
\end{itemize}
\item Would you behave differently during your real-life work?
\item Do you think your solutions would be compliant with data privacy? Why/why not?
\end{itemize}
\end{itemize}
If Prompted:
\begin{itemize}
\item How did you ensure data privacy compliance?
\begin{itemize}
\item Did you follow any guidelines or online sources?
\end{itemize}
\item What challenges did you face during the implementation w.r.t. privacy?
\item Did you use any online sources to help you with your challenges w.r.t. Privacy?
\begin{itemize}
\item What sources did you use?
\item How would you rate their effectiveness?
\item How do you feel they could be improved?
\end{itemize}
\item Would you behave differently during your real-life work?
\item Do you think your solutions would be compliant with data privacy? Why/why not?

\end{itemize}
If ExpertCommunication:
\begin{itemize}
\item How did you ensure data privacy compliance?
\begin{itemize}
\item Did you follow any guidelines or online sources?
\end{itemize}
\item What challenges did you face during the implementation w.r.t. privacy?
\item Did you use any online sources to help you with your challenges w.r.t. Privacy?
\begin{itemize}
\item What sources did you use?
\item How would you rate their effectiveness?
\item How do you feel they could be improved?
\end{itemize}
\item Would you behave differently during your real-life work?
\item Do you think your solutions would be compliant with data privacy? Why/why not?
\item You had the possibility to contact an expert during the implementation.
\begin{itemize}
\item Did you make use of this possibility? Why/Why not?
\item Was the support helpful? Why/Why not?
\item Would you like to have a similar privacy expert within your work context?
\end{itemize}
\end{itemize}
\begin{itemize}
\item Do you have any experience with similar privacy expert communication within your work context?
\item What could be improved w.r.t. the expert communication?
\end{itemize}
Concerning the backup fix:
\begin{itemize}
\item What do you think is required to provide a data privacy-compliant solution?
\item How have you/would you implement(ed) a data privacy-compliant solution?
\item What might help you with similar tasks in the future?
\end{itemize}
Concerning the Deletion Request:
\begin{itemize}
\item What do you think is required to provide a data privacy-compliant solution?
\item How have you/would you implement(ed) a data privacy-compliant solution?
\item What might help you with similar tasks in the future?
\end{itemize}
Concerning the Advertisement:
\begin{itemize}
\item What do you think is required to provide a data privacy-compliant solution?
\item How have you/would you implement(ed) a data privacy-compliant solution?
\item What might help you with similar tasks in the future?
\end{itemize}
Concerning the Search:
\begin{itemize}
\item What do you think is required to provide a data privacy-compliant solution?
\item How have you/would you implement(ed) a data privacy-compliant solution?
\item What might help you with similar tasks in the future?
\end{itemize}
This was my last question from my side, do you have anything else that you think is important that we have not talked about yet?

%% file: content/a3_screening.tex
\FloatBarrier
\section{Screening Survey}\label{apx:screening}

\noindent Please read the following consent form carefully. You will also be provided with a copy during
the lab study.

$\square$ I consent
$\square$ I do not consent\\

\label{Screening}

\noindent S1: You will have 90 seconds to answer the next question.\newline
\noindent Given the array arr[7,3,5,1,9], what could the command arr[3] return? The array index starts with 0.\newline
$\square$ 1 \newline
$\square$ 7, 3, 5\newline
$\square$ 21, 9, 15, 3, 27\newline
$\square$ 3 \newline
$\square$ 10, 6, 8, 4, 12 \newline
$\square$ I don't know \newline

\noindent S2: You will have 30 seconds to answer the next question.\newline
\noindent Which of these values would be the most fitting for a Boolean?\newline
$\square$ True \newline
$\square$ Red \newline
$\square$ Solid \newline
$\square$ Quadratic \newline
$\square$ Small \newline
$\square$ I don't know \newline

\noindent S3: You will have 60 seconds to answer the next question.\newline
\noindent Please select 2 items from the list which are IDEs.\newline
$\square$ Visual Studio \newline
$\square$ Eclipse \newline
$\square$ gcc \newline
$\square$ HTML\newline
$\square$ pyCharm \newline
$\square$ CSS \newline
$\square$ jpg \newline
$\square$ C++ \newline
$\square$ UDP \newline
$\square$ None of the above \newline
$\square$ I don't know \newline

\noindent S4: You will have 30 seconds to answer the next question.\newline
\noindent Choose the answer that best fits the definition of a recursive function.\newline
$\square$ A function that calls itself\newline
$\square$ A function that does not have a return value\newline
$\square$ A function that can be called from other functions\newline
$\square$ A function that runs for an infinite time \newline
$\square$ A function that does not require any inputs \newline
$\square$ I don't know \newline

\noindent S5: You will have 30 seconds to answer the next question.\newline
\noindent Which of these websites is used most frequently by developers as aid for programming?\newline
$\square$ Stack Overflow \newline
$\square$ Wikipedia \newline
$\square$ MemoryAlpha \newline
$\square$ LinkedIn \newline
$\square$ None of the above\newline

\noindent I1: Are you 18 years old or older?

$\square$ Yes
$\square$ No 

\noindent I2: Are you currently employed in IT or have been employed in IT in the last five years?

$\square$ No
$\square$ Yes 

\noindent I3:  How familiar are you with JavaScript?

\noindent$\square$ Not familiar at all \newline
$\square$ Slightly familiar \newline
$\square$ Moderately familiar \newline
$\square$ Very familiar \newline
$\square$ Extremely familiar \newline

%% file: content/a4_postsurvey.tex
\section{Post Survey}
\label{apx:post}
\noindent Thank you again for participating in our study! Please fill out this Survey only after you have finished the programming tasks. We will collect demographic information, as well as your thoughts on the study in this survey. Afterwards, we will conduct a short exit interview with you.\newline

\noindent Q1: How knowledgeable would you say when it comes to the General Data Protection
Regulation (GDPR)?\newline
$\square$ Not knowledgeable at all \newline
$\square$ Slightly knowledgeable\newline
$\square$ Moderately knowledgeable \newline
$\square$ Very knowledgeable \newline
$\square$ Extremely knowledgeable \newline

\noindent Q2: How often are tasks at you place of work deal with GDPR relevant issues?\newline
$\square$ Never \newline
$\square$ Some tasks \newline
$\square$ About half the tasks \newline
$\square$ Most of the tasks \newline
$\square$ All tasks \newline

 \noindent Q3: How confident are you that your solutions would be compliant with the General Data
Protection Regulation (GDPR)?\newline
$\square$ Not confident at all \newline
$\square$ Slightly confident \newline
$\square$ Moderately confident \newline
$\square$ Very confident \newline
$\square$ Extremely confident \newline

\noindent Q4: How old are you? (Text field)\newline

\noindent Q5: What is your gender?\newline
$\square$ Male \newline
$\square$ Female \newline
$\square$ Non-binary / third gender \newline
$\square$ Prefer not to say \newline
$\square$ Prefer to self-describe: (Text field)\newline

\noindent Q6: What is the highest level of school you have completed or the highest degree you have received?\newline
$\square$ Less than high school / GCSE or equivalent \newline
$\square$ High school or equivalent / A level or equivalent \newline
$\square$ Some college, currently enrolled in college, or two-year associate's degree, completed part of a higher education course, or currently enrolled \newline
$\square$ Vocational degree \newline
$\square$ Bachelor's degree \newline
$\square$ Some graduate school, or currently enrolled in graduate school \newline
$\square$ Master's or professional degree \newline
$\square$ Doctorate degree \newline
$\square$ Prefer not to say \newline
$\square$ Other: (Text field)\newline

\noindent Q7:  In which country do you currently reside?\newline
(List of 1357 countries from Afghanistan to Zimbabwe)\newline

\noindent Q8: What is your current employment status?\newline
$\square$ Employed full-time \newline
$\square$ Employed part-time \newline
$\square$ Prefer not to say \newline
$\square$ Other: (Text field) \newline

\noindent Q9: What is your current job title? (Text field) \newline

\noindent Q10: How many years have you been in the software industry? (Text field) \newline

\noindent Q11: How many people were employed in the largest company you worked for? (Approximate) (Text field) \newline

\noindent Q12: How many years of experience do you have in your current field? (e.g., 5 years as a software developer) (Text field) \newline

%% file: content/a10_metareview.tex
\newpage 


\section{Meta-Review}

The following meta-review was prepared by the program committee for the 2025
IEEE Symposium on Security and Privacy (S\&P) as part of the review process as
detailed in the call for papers.

\subsection{Summary}
This paper explores why developers struggle with building privacy-compliant implementations through a programming study with 30 professional software developers, in which they completed 1 warm-up task and 3 privacy-sensitive programming tasks. Follow-up interviews were then conducted. The results highlight a lack of privacy awareness further exacerbated by a perception that privacy requirements are complex and difficult to implement and that asking privacy experts for help will burden the experts. Participants lacked knowledge and experience, and as such, relied heavily on existing privacy implementations.

\subsection{Scientific Contributions}
\begin{itemize}
\item Provides a Valuable Step Forward in an Established Field
\item Creates a New Tool to Enable Future Science
\item Independent Confirmation of Important Results with Limited Prior Research
\end{itemize}

\subsection{Reasons for Acceptance}
\begin{enumerate}
\item This paper has a very strong methodology section, and it is very clear that there were careful considerations taken when designing this study to ensure the generalizability of results. The use of a lab study to avoid participants using AI in the study was a careful consideration.
\item This paper explores an understudied and important topic of privacy implications when developing software. Historically, the focus has been on security, so this paper offers a valuable foundation for future work.
\end{enumerate}